\documentclass[12pt]{article}

\makeatletter

\def\@authoraddress{}
\def\@title{}
\def\title#1{\gdef\@title{{\par\vskip-10pt\Large\bf
\baselineskip20pt\centering\ignorespaces\uppercase{#1}\vskip6pt}}%
\setcounter{table}{0}      \setcounter{figure}{0}
\setcounter{equation}{0}   \setcounter{section}{0}
\setcounter{subsection}{0} \setcounter{subsubsection}{0}
\setcounter{paragraph}{0}
}

\def\authors#1{\expandafter\def\expandafter\@authoraddress\expandafter
{\@authoraddress %
{\dimen0=-\prevdepth \advance\dimen0 by1.5\baselineskip
\nointerlineskip \centering
\vrule height\dimen0 width0pt\relax\ignorespaces\large\sc#1\par
}%
}%
}

\def\addresses#1{\expandafter\def\expandafter\@authoraddress\expandafter
{\@authoraddress{\nointerlineskip\vskip1pc
                 \footnotesize\it\centering\ignorespaces#1\par}}}

\def\nextaddress{\\[2.3pt]}

\def\@maketitle{%
\@title
\ifdim\prevdepth=-1000pt \prevdepth0pt\fi
\@authoraddress
}

\def\maketitle{\par
\begingroup
\let\cite\@bylinecite
\global\@topnum\z@ %
\@maketitle
\endgroup
\def\@thanks{}\def\@authoraddress{}\def\@title{}
}

\def\abstract{\par
\bgroup
\ifdim\prevdepth=-1000pt \prevdepth0pt\fi
\hsize\columnwidth
\leftskip=2em \rightskip\leftskip
\dimen0=-\prevdepth \advance\dimen0 by2pc \nointerlineskip
\noindent\vskip1.5\baselineskip\nointerlineskip\noindent\footnotesize\relax}

\newif\if@firststuff

\def\endabstract{\par
\nointerlineskip \vskip0pt
\noindent \par
\egroup
\hrule depth0pt width0pt
\global\everypar{\global\@firststufffalse}\global\@firststufftrue
}

\renewcommand\section{\@startsection {section}{1}{\z@}%
                                   {-3.5ex \@plus -1ex \@minus -.2ex}%
                                   {2.3ex \@plus.2ex}%
                                   {\normalfont\large\bfseries}}
\renewcommand\subsection{\@startsection{subsection}{2}{\z@}%
                                     {-3.25ex\@plus -1ex \@minus -.2ex}%
                                     {1.5ex \@plus .2ex}%
                                     {\normalfont\large\bfseries}}

\def\1ad{\mbox{\normalsize $^1$}}
\def\2ad{\mbox{\normalsize $^2$}}
\def\3ad{\mbox{\normalsize $^3$}}
\def\4ad{\mbox{\normalsize $^4$}}
\def\5ad{\mbox{\normalsize $^5$}}
\def\6ad{\mbox{\normalsize $^6$}}
\def\7ad{\mbox{\normalsize $^7$}}
\def\8ad{\mbox{\normalsize $^8$}}
\def\adref#1{\mbox{\normalsize $^{#1}$}}
\makeatother

\usepackage{amsfonts}
\usepackage{epsfig}

\begin{document}
\raggedbottom

\title{Phase transitions, massive gravitons and effective
action in braneworld theory}

\authors{A.O.~Barvinsky,\adref{1}
  A.Yu.~Kamenshchik,\adref{2,3}
  C.~Kiefer,\adref{4}
  and A.~Rathke\adref{4}}

\addresses{\1ad Theory Department, Lebedev Physics Institute,
Leninsky Pr. 53, Moscow 117924, Russia
  \nextaddress \2ad L.D. Landau Institute for Theoretical Physics of
Russian Academy of Sciences, Kosygina str. 2, Moscow 117334,
Russia
  \nextaddress \3ad Landau Network - Centro Volta, Villa Olmo, via Cantoni 1,
Como 22100, Italy
  \nextaddress \4ad Theoretical Physics Institute, University of Cologne,
Zuelpicher Str. 77, 50937 Cologne, Germany.}

\maketitle

\begin{abstract}
We construct the holographic type nonlocal effective action in
two-brane Randall-Sundrum model and show that it describes
a phase transition between the local and nonlocal phases of the
theory --- a cumulative effect of the tower of massive Kaluza-Klein
modes. We show that the corresponding renormalization group flow
interpolating between the limits of short and long interbrane
separations can be dynamically mediated by a repulsive interbrane
potential that gives rise to braneworld cosmological scenarios
with diverging branes.
\end{abstract}

\section{Introduction}
Recent developments in string theory
\cite{string} and the attempts to resolve the hierarchy problem
\cite{hierarchy} suggest that the observable world can be a brane
embedded in a higher-dimensional spacetime with a certain number
of noncompact dimensions. Moreover, string-inspired field theories
imply the existence of several branes interacting and propagating
in the multi-dimensional bulk. Their dynamics manifests itself for
the observer as an effective four-dimensional theory that, in the
cosmological context should explain the origin
of structure in the Universe by means of an inflationary or some
other scenario \cite{DTye,Ekpyr}, explain its particle
phenomenology, and shed light on problems such as a possibly
observable cosmological acceleration \cite{accel}.

The efficient way of description for the braneworld scenario is
the method of effective action. Here we calculate this action in
the two-brane Randall-Sundrum model within the holographic setup
characteristic of the AdS/CFT-correspondence interpretation of this
model \cite{Gubser,GKR,HHR1} -- the braneworld action as a
functional of two brane
geometries effectively incorporating the dynamics of multi-dimensional
gravitational field in the bulk. First we dwell on the definition
of this action as an alternative to the conventional Kaluza-Klein
description. In the approximation quadratic in the curvature we
obtain it as a nonlocal functional of two
induced metrics on branes and the corresponding radion fields. We show
how conventional Kaluza-Klein bulk modes arise as the set of zeros of
the nonlocal form factors of the action and analyze the first two
levels of the Kaluza-Klein tower -- massless and massive gravitons --
within the low-derivative gradient expansion. Then we consider the
reduced version of this action obtained by integrating out the fields
on the negative-tension brane invisible from the viewpoint of the
Planckian brane observer.  We show that for small interbrane separation
this action describes a Brans-Dicke type theory with the nonminimally
coupled radion field and {\em local} Weyl-squared short distance
corrections. For large interbrane distances it explicitly features
the recovery of the Einstein theory with {\em nonlocal} short-distance
corrections reflecting the well-known AdS/CFT correspondence
principle. The renormalization group flow responsible for this
transition between the local and nonlocal phases of the theory
can be realized dynamically by means of the repulsive interbrane
potential of the radion field. This generates the inflationary
\cite{brane} and other cosmological scenarios alternative to
models of colliding branes \cite{DTye,Ekpyr}.

\section{Braneworld effective action --
alternative to Kaluza-Klein reduction} \label{sec:1}

Let us clarify the difference between the Kaluza-Klein and
braneworld definitions of the effective actions. In Kaluza-Klein
setting the construction of the effective action consists in the
well-known procedure of decomposing the multi-dimensional field
$\Phi(x,y)$, where $x$ are the visible (four-dimensional)
coordinates and $y$ are the coordinates of extra dimensions, in a
certain complete set of harmonics $Z_n(y)$ on the $y$-space,
$\Phi(x,y)=\sum_n\phi_n(x)Z_n(y)$, and substituting the result
into the fundamental action $S[\,\Phi(x,y)\,]$ of the field
$\Phi(x,y)$. Subsequent integration over $y$ gives the effective
action for an infinite tower of fields $\{\phi_n(x)\}$ in
$x$-spacetime,
    \begin{equation}
    S[\,\Phi(x,y)\,]=
    \int dx\,dy\,L(\Phi(x,y),\partial\Phi(x,y))=
    \int dx\,L_{\rm eff}
    (\{\phi_n(x)\},\{\partial\phi_n(x)\}).  \label{1}
\end{equation}
The harmonics $Z_n(x)$ are usually taken as eigenmodes of the
$y$-part of the full wave operator of the theory with eigenvalues
$m_n^2$ -- the masses of the Kaluza-Klein modes. This type of action
was built for the two-brane Randall-Sundrum scenario
in \cite{KubVol}.  This  action, however, is very often not
helpful in the braneworld context, because it does not convey a
number of its important features like non-compactness of extra
dimensions \cite{RSloc,Rub}, recovery of the four-dimensional
Einstein gravity \cite{GT}, and its interpretation in terms of the
AdS/CFT-correspondence \cite{Gubser,GKR}.

These features can be described in a holographic formalism in which
a natural variable is the value of the field at the brane in question,
$\phi(x)=\Phi(x,y_{\rm brane})$. Unlike in Kaluza-Klein description,
its effective action $S_{\rm eff}[\,\phi(x)\,]$ is obtained from
the fundamental action $S[\,\Phi\,]$ by a more nontrivial
procedure -- by substituting in $S[\,\Phi\,]$ a solution of the
classical equations of motion for $\Phi(x,y)$ in the bulk,
$\Phi=\Phi[\,\phi(x)\,]$, parametrized by their boundary values on
the branes, that is $S_{\rm eff}[\,\phi(x)\,]
=S\big[\,\Phi[\phi(x)\,]\,\big]$. This construction obviously
generalizes to the case of several branes $\Sigma_I$ enumerated by
the index $I$ and the set of brane fields $\phi^I=\Phi(\Sigma_I)$.

Such a definition corresponds to the tree-level approximation for
the quantum effective action
    \begin{equation}
    \exp\Big(iS_{\rm eff}[\,\phi\,]\Big)=
    \left.\int D\Phi \,\exp\Big(iS[\,\Phi\,]\Big)\,
    \right|_{\,\Phi(\Sigma)\,=\,\phi}\,\, ,         \label{00.2}
    \end{equation}
where the functional integration over the bulk fields runs subject
to these brane boundary conditions. The scope of this formula is
very large, because it arises in very different contexts. In
particular, its Euclidean version $(iS\to -S_{\rm Euclid})$
underlies the construction of the no-boundary wavefunction in
quantum cosmology \cite{HH}. Semiclassically, in the braneworld
scenario, it represents a Hamilton-Jacobi functional, and its
evolutionary equations of motion in the ``fifth time'' $y$ can be
interpreted as renormalization-group equations \cite{Verlinde}. It
also underlies the effective action formulation of the
AdS/CFT-correspondence principle between supergravity theory on an
$AdS_5\times S^5$ background and the superconformal field theory
(super-Yang-Mills) on its infinitely remote boundary
\cite{AdS/CFT,GKR,HHR1}.

\section{Two-field action in two-brane Randall-Sundrum
model and massive gravitons} \label{sec:2} Here we consider the
two-brane Randall-Sundrum model with $Z_2$ orbifold identification
of points on the compactification circle of the fifth coordinate
\cite{RSloc}. The action of this model equals
     \begin{eqnarray}
     &&S[\,G,g\,]=\frac1{16\pi G_5}
     \int d^5x\,G^{1/2}
     \left(\,\vphantom{I}^5\!R(G)-2\Lambda_5\right)
     \nonumber\\
     &&\qquad\qquad\qquad\qquad
     +\sum\limits_{I}\int_{\Sigma_I}\!
     d^4x\,g^{1/2}\left(
     \frac1{8\pi G_5}[K]-\sigma_I\right), \label{0.1}
     \end{eqnarray}
where the index $I=\pm$ enumerates two branes with two brane
tensions $\sigma_\pm$ and $[K]$ is the trace of the extrinsic
curvature jump on branes. These branes are located
at antipoidal points of the circle labelled by the values of the
$y$, $y=y_\pm,\,y_+=0,\,|y_-|=d$. $Z_2$-symmetry identifies the
points on the circle $y$ and $-y$. When the brane tensions are
opposite in signs and fine tuned in magnitude to the values of the
negative cosmological constant $\Lambda_5$ and the 5-dimensional
gravitational constant $G_5$ according to the relations
     \begin{eqnarray}
     &&\Lambda_5=-\frac6{l^2},\,\,
     \sigma_+=-\sigma_-=\frac3{4\pi G_5l},   \label{3.2}
     \end{eqnarray}
then in the absence of matter on branes this model admits the
solution with the AdS metric in the bulk ($l$ is its curvature
radius),
     \begin{eqnarray}
     &&ds^2=dy^2+e^{-2|y|/l}\eta_{\mu\nu}dx^\mu dx^\nu,
     \end{eqnarray}
$0=y_+\leq|y|\leq y_-=d$, and with flat induced metric
$\eta_{\mu\nu}$ on both branes \cite{RSloc}. The metric on the
negative tension brane is rescaled by the value of
compactification factor $\exp(-2d/l)$ providing a possible
solution for the hierarchy problem \cite{RShier}. With fine tuning
(\ref{3.2}) this solution exists for arbitrary brane separation
$d$ -- two flat branes stay in equilibrium.

Take now the case when induced metrics on branes
differ from the background values by perturbations
    \begin{eqnarray}
    g^\pm_{\mu\nu}(x)=
    a^2_\pm\,\eta_{\mu\nu}+h^\pm_{\mu\nu}(x)  \label{1.4}
    \end{eqnarray}
(with $a_\pm=a(y_\pm)$ given in terms of the interbrane
distance $a_+=1,\,\,\,a_-=e^{-2d/l}\equiv a$), which induce
the perturbed solution of Einstein equations in the bulk
     \begin{eqnarray}
     ds^2=dy^2+e^{-2|y|/l}\eta_{\mu\nu}dx^\mu dx^\nu
     +h_{AB}(x,y)\,dx^Adx^B,                        \label{1.3}
     \end{eqnarray}
and consider the calculation of the braneworld action of the above
type in the approximation quadratic in $h^\pm_{\mu\nu}(x)$. In the
wording of Sect.2 the braneworld action of
$\phi(x)=(g_{\mu\nu}^\pm(x),\psi^\pm(x)$,
$S_{\rm eff}[\,\phi(x)\,]=S_4[\,g_{\mu\nu}^\pm(x),\psi^\pm(x)\,]$, is
invariant under the two four-dimensional diffeomorphisms
acting on the branes. In the linearized approximation they reduce
to the transformations of metric perturbations,
   $h^\pm_{\mu\nu}\rightarrow h^\pm_{\mu\nu}
   +f^\pm_{\mu\,,\,\nu}+f^\pm_{\nu\,,\,\mu}$,
with two {\em independent} local vector field parameters
$f_\mu^\pm=f_\mu^\pm(x)$. Therefore the action is expressible in
terms of the tensor invariants of these transformations ---
linearized Ricci tensors of $h_{\mu\nu}=h^\pm_{\mu\nu}(x)$,
  \begin{eqnarray}
  R^\pm_{\mu\nu}=\frac12\left(-\Box h_{\mu\nu}
  +h^\lambda_{\nu,\lambda\mu}
  +h^\lambda_{\mu,\lambda\nu}-h_{,\mu\nu}\right)^\pm,  \label{1.8}
  \end{eqnarray}
on {\em flat} four-dimensional backgrounds of both branes.

Due to metric perturbations the branes no longer stay at fixed
values of the fifth coordinate. Up to four-dimensional
diffeomorphisms, their embedding variables consist of
two four-dimensional scalars --- the radions $\psi^\pm(x)$ ---
which also enter as arguments of the braneworld action.
In the Randall-Sundrum gauge, $h_{A5}=0$,
${h_{\mu\nu}}^{,\,\nu}=h_\mu^\mu=0$, these embeddings are
defined by the equations
   \begin{eqnarray}
   \Sigma_\pm:\,\,\,
   y=y_\pm+\frac l{a^2_\pm}\psi^\pm(x),\,\,\,
   y_+=0,\,y_-=d.                          \label{1.11}
   \end{eqnarray}

The answer for the braneworld action, which we advocate
here, and which can be derived either by the direct substitution
of the linearized solution (\ref{1.3}) of bulk Einstein equations
in (\ref{0.1}) \cite{duality} or by functionally integrating
effective 4-dimensional equations of motion \cite{BWEA}, is
given in terms of the invariant fields of the above type,
$(R^\pm_{\mu\nu}(x),\psi^\pm(x))$, by the following spacetime
integral of a $2\times2$ quadratic form,
   \begin{eqnarray}
   &&S_4\,[\,g^\pm_{\mu\nu},\psi^\pm]
   =\frac1{16\pi G_4}\int d^4x \left[\,{\bf R}^T_{\mu\nu}\,
   \frac{2{\bf F}(\Box)}{l^2\Box^2}\,{\bf R}^{\mu\nu}
   +\frac16\,{\bf R}^{T}\,
   \frac{{\bf K}(\Box)\!-\!6{\bf F}(\Box)}{l^2\Box^2}
   \,{\bf R}\right.\nonumber\\
   &&\qquad\qquad\qquad\qquad\qquad
   \left.-3\Big(\Box{\bf\Psi}\!+\!\frac16{\bf R}\Big)^{\!T}\,
   \frac{{\bf K}(\Box)}{l^2\Box^2}\,
   \Big(\Box{\bf\Psi}\!+\!
   \frac16{\bf R}\Big)\,\right].      \label{1.12}
   \end{eqnarray}
Here $G_4$ is an effective four-dimensional gravitational coupling
constant,$ G_4=G_5/l$, $({\bf R}^{\mu\nu},{\bf\Psi})$ and $({\bf
R}^T_{\mu\nu},{\bf\Psi}^T)$ are the two-dimensional columns
  \begin{eqnarray}
   {\bf R}_{\mu\nu} =
   \left[\begin{array}{c}
      \,R_{\mu\nu}^+(x)\, \\ \,R_{\mu\nu}^-(x)\,
  \end{array}\right],\,\,\,\,\,
  {\bf\Psi}=\left[\begin{array}{c}
      \,\psi^+(x)\, \\ \, \psi^-(x)\,
  \end{array}\right]                       \label{1.13a}
   \end{eqnarray}
and rows ${\bf R}^T_{\mu\nu}=
   \Big[\,R^+_{\mu\nu}(x)\,\,\,
   R^-_{\mu\nu}(x)\,\Big],\,\,\,\,\,
   {\bf\Psi}^T=\Big[\,\psi^+(x)
   \,\,\,\psi^-(x)\,\Big]$
of the Ricci curvatures and radions associated with two branes. The nonlocal
form factors in (\ref{1.12}) express in terms of the nonlocal
operator ${\bf F}(\Box)$ the following $2\times2$-matrix valued
function of the $\Box$
    \begin{eqnarray}
    &&{\bf F}(\Box) =-\frac{1}{ J_2^+\,Y_2^- -J_2^-\,Y_2^+}
    \left[\begin{array}{cc}
    \,\sqrt{\Box} z_+u_+(z_-)&-2/\pi \\
    -2/\pi &\sqrt{\Box} z_-u_-(z_+)\,
    \end{array}\right],                            \label{4.11}\\
    &&u_\pm(z)=Y_1^\pm J_2(z\sqrt\Box)
    -J_1^\pm Y_2(z\sqrt\Box),                        \label{4.6}\\
    &&J_\nu^\pm\equiv J_\nu(z_\pm\sqrt\Box),\,\,\,
    Y_\nu^\pm\equiv Y_\nu(z_\pm\sqrt\Box),
    \end{eqnarray}
expressible in terms of the Bessel ($J_\nu$) and Neumann ($Y_\nu$)
cylindrical functions. The second form factor ${\bf K}(\Box)$
equals ${\bf K}(\Box)={\bf F}(\Box)+l^2\Box\,{\rm diag}[-1,\,1/a^2]$.

The zero-eigenvalue eigenvectors of these nonlocal operators
correspond to propagating modes of the theory. In the sector of
transverse-traceless metric perturbations (graviton sector), for
example, these modes ${\bf v}_n(x)$ are defined by the equation
${\bf F}(m_n^2){\bf v}_n=0$, where $m_n$ are the masses of these
excitations, $(\Box-m_n^2){\bf v}_n(x)=0$, given by the
roots $\Box=m_n^2$ of the equation ${\rm det}{\bf F}(\Box)=0$. For
the operator (\ref{4.11}) these equation reduces to
    \begin{equation}
    {\rm det}\,{\bf F}(\Box)\sim
    \Box\,\Big(Y_1^-J_1^+-Y_1^+J_1^-\Big)=0  \label{4.33}
    \end{equation}
and thus gives a well-known mass spectrum of Kaluza-Klein modes in the
Randall-Sundrum model, because these combination of Bessel functions
is exactly the left-hand side of the eigenvalue problem for the
harmonics $Z_n(y)$ satisfying the relevant Neumann-type boundary
conditions on the two branes.

Eq.~(\ref{4.33}) shows that the massless graviton is guaranteed to exist
in the spectrum, its localization on the positive-tension (Planckian)
brane reflecting the recovery of Einstein theory on this brane for
low energies satisfying the bounds
    \begin{equation}
    l\sqrt\Box\ll 1,\,\,\, \frac{l\sqrt\Box}a\ll 1.  \label{4.19}
    \end{equation}
In this domain the nonlocal form factor can be expanded in powers of
derivatives to have the structure
   \begin{eqnarray}
    &&\frac{\bf F(\Box)}{l^2}=
    -{\bf M}+{\bf D}\Box+O(\Box^2),    \label{4.26}\\
    &&{\bf M}\!=\!\frac1{l^2}\,\frac4{1-a^4}\,
    \left[\begin{array}{cc}
    \!a^4 & -a^2\!\! \\
    \!-a^2 & 1 \end{array}\right],\,
    {\bf D}\!=\!\frac{1-a^2}{6(1+a^2)^2}
    \left[\begin{array}{cc}
    \!a^2+3 & 2 \\
    2 & 3+a^{-2}\!\!
    \end{array}\right],                 \label{4.28}
    \end{eqnarray}
which contains both the mass matrix ${\bf M}$ and the
matrix of the kinetic term ${\bf D}$. The mass matrix is
degenerate in accordance with the presence of the massless
mode, ${\rm det}\,{\bf M}=0$, ${\rm rank}\,{\bf M}=1$, while
the kinetic matrix is positive definite. This allows one to
simultaneously diagonalize the both matrices in the basis of
new fields $h^0_{\mu\nu}$ and $h^M_{\mu\nu}$, so that the
low-energy action in the graviton sector describes two fields
--- the massless graviton and the massive transverse-traceless
tensor of the mass $M$,
    \begin{eqnarray}
    &&S_{\rm graviton}[\,h^0_{\mu\nu},h^M_{\mu\nu}]
    =\frac1{32\pi G_4}\int
    d^4x \Big(\,h_0\Box\,h_0
    +h_M(\Box-M^2)\,
    h_M\Big),                  \label{4.31}\\
    &&M^2=\frac{24}{l^2}\,
    \frac{a^2(1+a^2)}{(1-a^2)^2}.  \label{4.30}
    \end{eqnarray}
In view of standard arguments of gauge invariance under the
two diffeomorphism transformations above, the massless graviton has two
dynamical degrees of freedom, while the massive tensor field has
all five polarizations of a generic transverse-traceless tensor
field.

Strictly speaking, even the lowest-lying massive mode cannot be
consistently described within the low-derivative expansion (because
the mass (\ref{4.30}) as a candidate for $m_1^2$ violates the
second of the low-energy restrictions (\ref{4.19})). However,
numerical calculation of the first nontrivial root,
$m_1=\sqrt{\Box}$, of (\ref{4.33}) shows
that for a wide range of values of $a$ it coincides with good
precision with (\ref{4.30}). Therefore, the massive graviton
mode with the mass (\ref{4.30}) seems to provide a good description
of the first non-zero level. Its application in the form of the
phenomenon of radion induced graviton oscillations, potentially
interesting in the gravitational wave astronomy, is considered in
\cite{RIGO}.

\section{Local to nonlocal phase transitions and AdS/CFT
corre\-spondence} \label{sec:3}
The difficulties with a consistent description of massive modes
originate from the fact that the low-energy approximation on
the Planckian brane turns out to belong the physical high-energy
limit on the negative-tension brane (remember that the physical
energy there is determined by the scale of
$\eta^{\mu\nu}\partial_\mu\partial_\nu/a^2$). This situation becomes
even worse for large brane separation $a=\exp(-d/l)\to 0$. To avoid
this difficulty let us consider the reduced version of the braneworld
action corresponding to tracing (integrating) out the fields on the
negative-tension brane. This procedure can be motivated by a simple
physical fact that this brane is invisible from the viewpoint of the
observer sitting on the Planckian brane. In the tree-level
approximation this reduction, $S_{\rm eff}[\,\phi^+,\phi^-]
\Rightarrow S_{\rm red}[\,\phi^+]$, is equivalent to the exclusion of
the fields on the negative-tension brane in terms of those on the
positive-tension one, $S_{\rm red}[\,\phi^+]=
S_{\rm eff}\big[\,\phi^+,\phi^-[\,\phi^+]\big]$, as solutions
of their respective equations of motion,
$\delta S_{\rm eff}[\,\phi^\pm]/\delta\phi^-=0$,
$\phi^-=\phi^-[\,\phi^+]$.

Below we present the result of this reduction in two energy domains
--- one corresponding to (\ref{4.19}) and another for the case of
large brane separation when the second of inequalities (\ref{4.19})
is violated. For small or finite interbrane distance in the range of
(\ref{4.19}) the low-energy reduced action has the following form
     \begin{eqnarray}
     &&S_{\rm red}[\,g_{\mu\nu},\varphi\,]=\int d^4x \sqrt{g}
     \left[\left(\frac{m_P^2}{16\pi}
     -\frac1{12} \varphi^2\right)R
     +\frac12\varphi\,\Box\varphi\right. \nonumber\\
     &&\qquad\qquad\qquad\qquad\qquad\qquad\quad\qquad\qquad
     \left.
     +\frac{l^2 m_P^2}{32\pi}\,
     \kappa(\varphi^2)\,
    C_{\mu\nu\alpha\beta}^2\,
     \right],                           \label{5.1}\\
     &&\kappa(\varphi^2)=\frac14\,
    \left[\,\ln\frac1{a^2}-(1-a^2)-
      \frac12 (1-a^2)^2\,
      \right]_{\,a^2=4\pi\varphi^2/3m_P^2}.       \label{8.10a}
     \end{eqnarray}
in terms of the new scalar field $\varphi(x)$ --- the reparametrization of
the original radion field,
    \begin{equation}
    \varphi(x)=\sqrt{\frac3{4\pi}}m_P\,e^{-d/l-\psi(x)},
    \end{equation}
and the four-dimensional Planck mass, $m_P=\sqrt{l/G_5}$. This action
represents the Eistein gravity nonminimally coupled to the Brans-Dicke type
scalar $\varphi$ describing the local, depending on $x$, distance between
the branes and having the short distance corrections in terms of
the squared Weyl tensor $C_{\mu\nu\alpha\beta}^2$ with local
(but $\varphi$-dependent) coeffficient (\ref{8.10a}).

For large interbrane distance corresponding to high-energy domain
on the invisible brane,
      \begin{equation}
      l\sqrt\Box\ll 1,\,\frac{\sqrt\Box}a\gg1,
      \end{equation}
this action has a different behavior
    \begin{eqnarray}
    &&S_{\rm red}[\,g_{\mu\nu}\,]
    =\frac{m_P^2}{16\pi}\int d^4x\,\sqrt{g}
    \left[\,R+\frac{l^2}2\,
      C_{\mu\nu\alpha\beta}\,k(\Box)
      C^{\mu\nu\alpha\beta}\,\right], \label{8.24}\\
    &&k(\Box)=
    \frac14\left(\,\ln\frac4{l^2(-\Box)}
    -{\bf C}\,\right).                    \label{8.26}
    \end{eqnarray}
Radion field essentially decouples from gravity and Weyl-squared term
becomes nonlocal with the logarithmic form factor characteristic of
the AdS/CFT-correspondence phenomenon --- immitation of quantum
logarithms of conformal field theory on the brane by the tree-level
(super)gravitational action calculated in the bulk
\cite{Gubser,GKR,HHR1}.

Transition from the local phase (\ref{5.1}) to the nonlocal phase
(\ref{8.24}) of the theory represents the renormalization group flow
(AdS flow) interpolating between the limits of small and large
interbrane distances. The scalar field $\varphi$ starting at
$\varphi=\sqrt{3/4\pi}m_P$, ($a=1$ -- the point of coinciding branes)
tends to zero, ($a\to 0$ -- the limit of large interbrane separation).
The condensate of this field in the form of the
Coleman-Weinberg type of potential logarithmic in $\varphi^2/m_P^2$,
$\kappa(\varphi^2)$, (\ref{8.10a}), then delocalizes into the logarithmic
form factor $k(\Box)$, (\ref{8.26}), of the Weyl squared term. The
dominant logarithmic term of $\kappa(\varphi^2)
\sim (1/4)\ln(m_P^2/\varphi^2)$ at $\varphi\to 0$ instead of an
infinite growth saturates
at the logarithmic scale of the graviton radiation characterized by
spacetime inhomogeneity of the Weyl tensor -- the logarithmic
nonlocality of $k(\Box)\sim(1/4)\ln(4/l^2\Box)$. The physics of
this transition is transparent: the tower of massive
Kaluza-Klein modes which are infinitely heavy at the initial point
of merging branes (see (\ref{4.30}) at $a\to 1$), for
$a\to 0$ becomes very light --- its spectrum getting practically
continuous and resulting in a cumulative effect in the form of the
logarithmic nonlocality characteristic of the AdS/CFT-correspondence.

\section{Conclusions. Radion induced inflation and dynamical
realization of the AdS flow}
Thus, we have constructed the holographic type braneworld action
in the two-brane Randall-Sundrum model, demonstrated the existence
of massive graviton modes in its spectrum and showed that it features
the renormalization group flow from small to large interbrane
distances associated with different, respectively, local and nonlocal
phases of the theory.

From physics viewpoint it is certainly
interesting to see whether this flow can be realized at the dynamical
level, that is enforced as a solution of effective equations of
motion. In \cite{brane} it was suggested that a small detuning of
brane tensions from their Randall-Sundrum values (\ref{3.2}) leads to
the origin of the four-dimensional cosmological term in the action
(\ref{5.1}). Then after the reparametrization of this action to the
Einstein frame of new fields $(\bar g_{\mu\nu},\phi)$ (with $\phi$
minimally coupled to $\bar g_{\mu\nu}$ and related to the old field
$\varphi$ by
$\varphi=\sqrt{3/4\pi}\,m_P\tanh(\sqrt{4\pi/3}\,\phi/m_P)$) one
finds that the new field $\phi$ acquires the potential,
$V(\phi)\sim \cosh^4(\sqrt{4\pi/3}\,\phi/m_P)$, capable to
maintane a slow-roll inflation. It corresponds to
two branes diverging under the repulsive force of this interbrane
potential (remember, that the limit $\phi\to 0$ corresponds to
$a\to 0$) and, thus, dynamically realizes the renormalization flow
of the above type. This model qualitatively differs from the inflation
and other scenarios of colliding branes \cite{DTye,Ekpyr}.
Unfortunately, this model suffers from an essential drawback.
This is the necessity to introduce by hand a four-dimensional
cosmological constant --- the brane tension detuning of the above
type.

Interestingly, the present results suggest a natural mechanism
of interbrane repulsion based on the Weyl-squared term of
(\ref{5.1}) and (\ref{8.24}). When the brane Universe is
filled with the graviton radiation $C_{\mu\nu\alpha\beta}^2>0$,
and for small brane separation this term forms the interbrane
potential $-(l^2/2)\kappa(\varphi^2)C_{\mu\nu\alpha\beta}^2$. It has
a maximum
at the point of coinciding branes $a=1$ because the coefficient
$\kappa(\varphi^2)$, Eq.~(\ref{8.10a}), is strictly positive. The
repelling force is very small, though, and vanishes
at $a=1$, because of the behaviour of
$\kappa(\varphi^2)\sim(1-a^2)^3/12$, $a^2=4\pi\varphi^2/3m_P^2$
(note that the expression (\ref{8.10a})
represents  the logarithm $\ln(1/a^2)$ with exactly
the first two terms of its Taylor series in $(1-a^2)$
subtracted). Unfortunately, this
potential is strictly negative, because $\kappa(\varphi^2)\geq 0$, and,
therefore, cannot maintain
inflation. Rather, it can serve as a basis of brane
models with $AdS_4$ geometry embedded in $AdS_5$ \cite{AdS_4brane}.
It can also be useful in the model of ``thick'' branes in the
Big Crunch/Big Bang transitions of ekpyrotic and
cyclic cosmologies \cite{Ekpyr}.

Thus, the two-brane model is likely to have a natural
mechanism to realize the AdS flow dynamically. Concrete implications
of this flow and the corresponding transitions between the local
and nonlocal phases of the theory still have to be worked out.
In particular, they might be important in view of the growing
interest in nonlocal modifications of gravity within the
cosmological constant problem \cite{AHDDG}. They are currently
under study.

\section*{Acknowledgements}

A.O.B.\ and A.Yu.K.\ are grateful for the
hospitality of the Theoretical Physics Institute, University of Cologne, where
a major part of this work has been done due to the support of the DFG grant
436 RUS 113/333/0-2. A.Yu.K.\ is also grateful to CARIPLO Science Foundation.
The work of A.O.B.\ was supported by the Russian  Foundation for
Basic Research under the grants No 02-02-17054 and No 00-15-96566, while
A.Yu.K.\ was supported by the RFBR grants No 02-02-16817 and
No 00-15-96699. A.R.\ is supported by the DFG Graduiertenkolleg
``Nonlinear Differential Equations''.

\end{document}